\documentclass[twocolumn]{revtex4}

\usepackage{amsmath}
\usepackage{graphicx}
\usepackage{xspace}

\graphicspath{{./figures/}}

\newcommand{\superk}      {Super-Kamiokande\xspace}       

\newcommand{\nue}         {$\nu_{e}$\xspace}
\newcommand{\numu}        {$\nu_{\mu}$\xspace}
\newcommand{\nutau}       {$\nu_{\tau}$\xspace}

\newcommand{\dms}         {$\Delta m^2$\xspace}
\newcommand{\sstt}        {$\sin^2 2 \theta$\xspace}

\newcommand{\tonetwo}     {$\theta_{12}$\xspace}
\newcommand{\tonethree}   {$\theta_{13}$\xspace}
\newcommand{\ttwothree}   {$\theta_{23}$\xspace}

\newcommand{\SK}          {Super-K\xspace}

\newcommand{\nukp}    {$p \to \bar{\nu} K^+$\xspace}
\newcommand{\nukz}    {$n \to \bar{\nu} K^0$\xspace}
\newcommand{\mukz}    {$p \to \mu^+ K^0$\xspace}
\newcommand{\ekz}     {$p \to e^+ K^0$\xspace}

\begin{document}
\title[Super-K]{The \superk Experiment \footnote{Prepared for
    {\it Neutrino Oscillations: Present Status and Future Plans},
    J. Thomas and P. Vahle editors, World Scientific Publishing
    Company, 2008.}  }

\author{Christopher W. Walter\footnote{For the \superk
    collaboration.} }
%\aindx{Walter, C.W.}

\address{Department of Physics, Duke University, Durham, NC 27708 USA \\
  chris.walter@duke.edu}

\begin{abstract}
  Super-Kamiokande is a 50~kiloton water Cherenkov detector located at
  the Kamioka Observatory of the Institute for Cosmic Ray Research,
  University of Tokyo.  It was designed to study neutrino oscillations
  and carry out searches for the decay of the nucleon.  The \superk
  experiment began in 1996 and in the ensuing decade of running has
  produced extremely important results in the fields of atmospheric
  and solar neutrino oscillations, along with setting stringent limits
  on the decay of the nucleon and the existence of dark matter and
  astrophysical sources of neutrinos.  Perhaps most crucially, \superk
  for the first time definitively showed that neutrinos have mass and
  undergo flavor oscillations. This chapter will summarize the
  published scientific output of the experiment with a particular
  emphasis on the atmospheric neutrino results.
\end{abstract}

\maketitle
%\body

\section{Introduction and Physics Goals}

The \SK collaboration is the combination of two previous successful
collaborations.  The first was the
Kamiokande~\cite{Hirata:1988uy, Hirata:1992ku} experiment which was
located in the same mine as \SK and had a fiducial mass approximately
20 times smaller than \SK.  The second was the IMB
experiment~\citep{Casper:1991ac, Becker-Szendy:1992hq}  which was
located in the Morton Salt mine in Ohio.  Grand Unified models such as
SU(5)~\cite{Georgi:1974sy} predicted that the proton would decay at a
rate visible by modest size detectors and both of these experiments
were originally built to search nucleon decay into the mode favored by
SU(5) which is $p \rightarrow e^+ \pi^0$.

Although neither of these experiments observed the decay of the
proton, they did measure a statistically significant anomaly in the
expected background to the proton decay search from neutrino
interactions on the water in the tanks.  One explanation for this
effect was that some of the neutrinos were oscillating into an
un-observable flavor and thus giving less background than expected.
At the same time two detectors made of iron the
NUSEX~\cite{Aglietta:1988be} and Frejus~\cite{Daum:1994bf} experiments
saw a result which was consistent with the expectation but with much
lower statistics.  \superk was designed to definitively determine
whether or not oscillations were indeed taking place.

Additionally, by scaling up the size from previous detectors, \superk
offered new hope to finally observe the decay of the nucleon and
also to try to answer the crucial question of whether neutrinos
produced in the nuclear burning in the sun oscillated into
non-detectable flavors on their way to the earth. Previous generations
of experiments had not seen as many neutrinos from the sun as
predicted by solar models.  With a large mass, good energy resolution,
and the ability to point neutrinos back to the sun in real-time, \superk
was designed first of all to confirm the effect with high statistics
and then to determine what the parameters of oscillation were.

\section{The Super-Kamiokande Detector}

% \begin{verbatim}
% SK [4]
% -----
% NIM      "{The Super-Kamiokande detector}",
% N16      "{N-16 as a calibration source for Super-Kamiokande}",
% Radon    "{Measurement of radon concentrations at Super-Kamiokande}",
% linac    "{Calibration of Super-Kamiokande using an electron linac}",
% \end{verbatim}

\begin{figure}[!b]
  \begin{center}
    \includegraphics[width=3.20in]{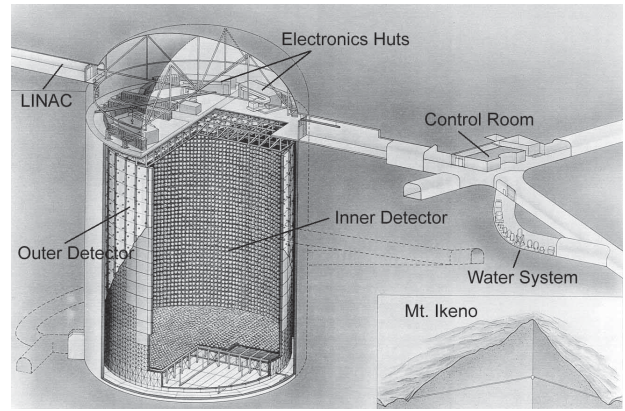}
    \caption{An overview of the \superk detector site, under Mt. Ikeno
      from Ref.~\cite{fukuda:2002uc}.  The cutaway shows the inside
      lined with photomultiplier tubes comprising a photo-cathode
      coverage of about 40\%. }
    \label{fig:detector}
  \end{center}
\end{figure}

Super-Kamiokande is a 50~kiloton water Cherenkov detector located at
the Kamioka Observatory of the Institute for Cosmic Ray Research,
University of Tokyo. Figure~\ref{fig:detector} shows a 
diagram of the Super-Kamiokande detector. The detector is in the
Mozumi mine of the Kamioka Mining Company in Gifu prefecture, in the
Japanese alps.  Super-K consists of two concentric, optically
separated water Cherenkov detectors contained in a stainless steel
tank 42 meters high and 39.3 meters in diameter, holding a total mass
of 50,000 tons of water.  The inner detector is comprised of
11,146 Hamamatsu~R3600 50~cm diameter photomultiplier tubes,
viewing a cylindrical volume of pure water 16.9~m in radius and 36.2~m
high. 

As described more fully in Ref.~\cite{fukuda:2002uc}, the detector
is calibrated in energy at the 2\% level over enegies ranging from the
MeV to the tens of GeV range.  This careful calibration is key to the
successful extraction of physics. It relies both on natural
calibration sources such as the expected energy deposit of muons
created by the cosmic rays and the decay of neutral pizeros produced
by neutrino interactions inside the \superk tank, and on artificial
sources including lasers, a Xenon light source, a linear
accelerator~\cite{nakahata:1998pz} and a ${}^{16}N$
source~\cite{blaufuss:2000tp}. Low energy radioactive background were
also carefully studied, a description of the measurement of radon
concentrations at \superk is given in Ref.~\cite{takeuchi:1999zq}.

Data from the detector is first collected by an on-line data
acquisition system and then, after a calibration step, passed into
several streams of reduction for the various analyses.  Selection
steps are performed to remove the background from non-neutrino induced
interactions.  For example, the outer detector region is used as a
veto to reject cosmic-ray muons.  Using the time and charge
information at each photo-tube, reconstruction algorithms are applied
to the data to determine a vertex for the Cherenkov light and any
rings associated with the particles of the interaction.  Additional
likelihood-based algorithms are used to determine the properties of
the particles that generated the light including their type, momenta
and directions.  These reconstructed physics quantities are then used
for analysis.  More detail about the acquisition and analysis
reconstruction techniques can be found in
Refs.~\cite{fukuda:2002uc, ashie:2005ik}
and~\cite{Hosaka:2005um}.
 
The \superk running periods are divided into three parts.  The first,
SK-I ran from 1996 to 2001.  In November of 2001 an accident destroyed
much of the photo-tubes of Super-K and, after a year of rebuilding 
the detector with half of the previous photo-tube density, SK-II ran for
approximately 800 days.  In June of 2006 after restoring \superk to
its full photo-tube density a period of running known as SK-III
began.  During the SK-I and SK-II running period \superk acted as the
target for the long-baseline K2K experiment.  Starting in
2009, \superk will once again be the target of an accelerator produced
neutrino beam when the T2K experiment begins.

\section{Published results from Super-Kamiokande}

The published scientific output of the \superk experiment can be
roughly divided into four main categories:

\begin{enumerate}
\item Studies of atmospheric neutrino oscillations
\item Studies of solar neutrino oscillations
\item Searches for the decay of the nucleon
\item Searches for astrophysical sources of neutrinos
\end{enumerate}

In the sections that follow, papers from all of these subjects will be
reviewed.  Particular attention will be payed to the history of the
published papers in the atmospheric neutrino analysis.

\subsection{Atmospheric Neutrino Oscillations}

{\bf The sub-GeV R ratio~\cite{fukuda:1998tw}}

In the previous results from the IMB and Kamiokande experiment it was
observed that the flavor ratio of neutrinos below 1~GeV
% \begin{equation}
%   (\nu_\mu+\overline{\nu}_\mu)/(\nu_e+\overline{\nu}_e),
% \end{equation}
% \noindent
did not agree with expectation.  If one took the results at face value,
either there were more electron neutrinos then expected, or too few
muon neutrinos.  In order to study the question experimentally a
measurement was made of:

\begin{equation}
R \equiv (\mu/e)_{DATA}/(\mu/e)_{MC},
\label{eqn:R}
\end{equation}

\noindent
which served to cancel uncertainties in neutrino flux and
cross-sections.  In Eqn.~\ref{eqn:R}, $(\mu/e)$ means the ratio of the
number of measured neutrino interactions which are inferred to have
come from electron and muon neutrinos respectively.  The ratio is
calculated separately for the reconstructed data and MC.  If there is
perfect agreement between data and expectation the expected value of
$R$ is one.  In \superk, the measured value of R was:
\begin{equation}
      R  =   0.61 \pm 0.03(stat.) \pm 0.05(sys.).
\end{equation}

This was a statistically significant result and the collaboration
concluded in Ref.~\cite{fukuda:1998tw}:

{\it ``The first measurements of atmospheric neutrinos in the
Super-Kamiokande experiment have confirmed the existence of a smaller
atmospheric $\nu_\mu/\nu_e$ ratio than predicted.  We
obtained $R = 0.61 \pm 0.03(stat.) \pm 0.05(sys.)$ for events in
the sub-GeV range.  The Super-Kamiokande detector has much greater
fiducial mass and sensitivity than prior experiments.  Given the
relative certainty in this result, statistical fluctuations can no
longer explain the deviation of $R$ from unity.''} \\

\noindent
{\bf The multi-GeV R ratio~\cite{fukuda:1998ub}}

The previous result relied on events which had visible energy less
that 1.33~GeV deposited in the \SK tank (so called sub-GeV events).
Although less numerous, multi-GeV events were also expected to
oscillate and had the extra advantage that at high-energy the outgoing
lepton direction closely followed the incoming neutrino direction.
Since the neutrino oscillation probability is a function of both the
distance traveled and the energy, knowing the direction of the
incoming neutrino allowed the separation of neutrinos into bins of
angular zenith.  Based on the results of the previous experiments it
was expected that neutrinos of a few GeV would need to travel
thousands of kilometers before they oscillated.  Also, if muon
neutrino were oscillating into tau neutrinos, the vast majority of
them would not have the energy necessary to interact and produce a tau
lepton.  In this case, the expectation was that there would be a
deficit of muon interactions coming from below.

From the data analyzed, the $R$ ratio for the multi-GeV events was 
reported in Ref.~\cite{fukuda:1998ub} to be:
\begin{equation}
  R  = 0.66 \pm 0.06(stat.) \pm 0.08(sys.).
\end{equation}

\noindent confirming the result seen in the sub-GeV sample.
Crucially, the expected zenith suppression was also seen. \\

\noindent
{\bf Evidence for oscillations~\cite{fukuda:1998mi}}

\begin{figure}[!htb]
  \begin{center}
    \includegraphics[width=3.0in]{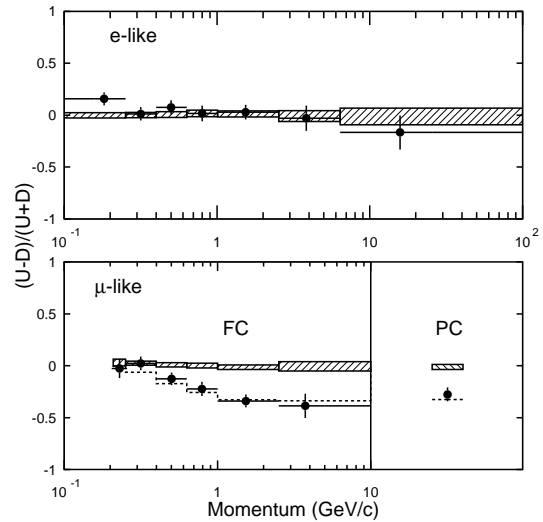}
    \caption{The $(U-D)/(U+D)$ asymmetry as a function of momentum for
      FC $e$-like and $\mu$-like events and PC events. While it is not
      possible to assign a momentum to a PC event, the PC sample is
      estimated to have a mean neutrino energy of 15 GeV. The Monte
      Carlo expectation without neutrino oscillations is shown in the
      hatched region with statistical and systematic errors added in
      quadrature. The dashed line for $\mu$-like is the expectation
      for $\nu_\mu \leftrightarrow \nu_\tau$ oscillations with
      (\sstt=1.0, \dms = 2.2 $\times 10^{-3}$ eV$^2$). Figure and
      caption from Ref.~\cite{fukuda:1998mi}. }
    \label{fig:asym}
  \end{center}
\end{figure}

In 1998, the paper ``Evidence for oscillation of atmospheric
neutrinos'' was published\cite{fukuda:1998mi}.  In this paper, 535
days of data were analyzed and subdivided into sub-GeV and multi-GeV
e-like and mu-like events.  In this data set, a strong suppression of
upward going event from muon-neutrino interactions was observed.  This
was quantified in terms of an asymmetry defined as $A = (U-D)/(U+D)$
where $U$ is the number of upward-going events ($-1 < \cos \Theta <
-0.2$) and $D$ is the number of downward-going events ($0.2 < \cos
\Theta < 1$). Due to the isotropic nature of the cosmic rays this
asymmetry is expected to be close to zero.  For the electron-neutrino
sample this was found to be the case.  For the muon-like events the
asymmetry was found to be:

\begin{equation}
  A = {(U-D) \over (U+D)} = -0.296 \pm 0.048(stat.) \pm 0.01(sys.) 
\end{equation}

\noindent which deviates from zero by more than 6 standard deviations.
Figure~\ref{fig:asym} taken from Ref.~\cite{fukuda:1998mi} shows
this asymmetry as a function of momentum for both the e-like and
mu-like samples.  The deficit of high energy upward-going mu-like
events is clearly seen.

\begin{figure*}[!htb]
%  \begin{center}
    \includegraphics[width=6.5in]{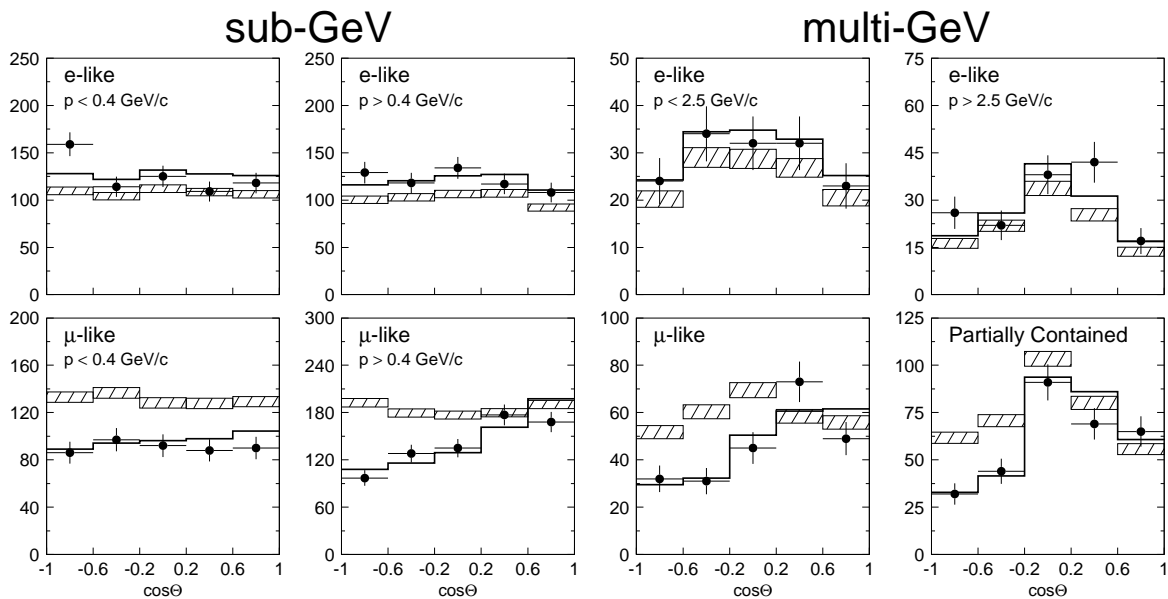}
    \caption{Zenith angle distributions of $\mu$-like and $e$-like
      events for sub-GeV and multi-GeV data sets. Upward-going
      particles have $\cos \Theta < 0$ and downward-going particles
      have $\cos \Theta > 0$.  Sub-GeV data are shown separately for
      $p < 400 $ MeV$/c$ and $p > 400 $ MeV$/c$. Multi-GeV $e$-like
      distributions are shown for $p < 2.5 $ GeV$/c$ and $p > 2.5 $
      GeV$/c$ and the multi-GeV $\mu$-like are shown separately for FC
      and PC events. The hatched region shows the Monte Carlo
      expectation for no oscillations normalized to the data live-time
      with statistical errors. The bold line is the best-fit
      expectation for \numu $\leftrightarrow$ \nutau oscillations with
      the overall flux normalization fitted as a free parameter.
      Figure and caption from Ref.~\cite{fukuda:1998mi}.} 
    \label{fig:angdist}
 % \end{center}
\end{figure*}

A fit was performed over the oscillation parameter space to the
$\nu_\mu \leftrightarrow \nu_\tau$ oscillation hypothesis. There were
eight systematic uncertainty terms that were allowed to vary within
their known ranges.  Fits were also performed to the $\nu_\mu
\leftrightarrow \nu_e$ hypothesis but the data did not fit this
hypothesis well.  The data-samples and the results of the fit are
shown in the Fig.~\ref{fig:angdist} also from
Ref.~\cite{fukuda:1998mi}.

This analysis resulted in the first allowed region for atmospheric
neutrino oscillations from \superk, resulting in a somewhat lower
allowed region then was previously obtained from the Kamiokande
collaboration.  In this paper the
conclusion was: \\

{\it Both the zenith angle distribution of $\mu$-like events and the value
of $R$ observed in this experiment significantly differ from the best
predictions in the absence of neutrino oscillations. While
uncertainties in the flux prediction, cross sections, and experimental
biases are ruled out as explanations of the observations, the present
data are in good agreement with two-flavor $\nu_\mu
\leftrightarrow \nu_\tau$ oscillations with \sstt$ > 0.82$ and
$5\times10^{-4} < $\dms$ < 6\times10^{-3}$ eV$^2$ at 90\% confidence
level. We conclude that the present data give
evidence for neutrino oscillations.} \\

It should be noted that the allowed region found in 1998 is consistent
with all of the later more refined analysis, which have also all been
consistent with each other. \\

\noindent
{\bf Upward-going~\cite{fukuda:1998ah} and
  stopping~\cite{fukuda:1999pp} muon samples}

Also analyzed separately in Refs~\cite{fukuda:1998ah} and
\cite{fukuda:1999pp} were upward through-going and stopping muon
samples.  Muons are created by neutrino interactions under the \SK
tank and then travel upwards, either passing through or stopping
inside the detector.  The through-going events were of a higher energy
than the events analyzed in Ref.~\cite{fukuda:1998mi} which were
completely contained inside the \SK tank, and the stopping events were
of comparable energy to the events previously considered in which the
produced muon escaped the tank.  The reduction process for these
events was somewhat more involved since it was more difficult to
remove backgrounds from entering downward-going muons.

In these two papers, the two sub-samples were fit to the oscillation
hypothesis as was their ratio as a function of zenith angle.  It was
found that they also gave results consistent with neutrino
oscillations as presented in Ref.~\cite{fukuda:1998mi}. \\

\noindent
{\bf The East-West~\cite{futagami:1999wz} effect}

\begin{figure}[!htb]
  \begin{center}
    \includegraphics[width=3.50in]{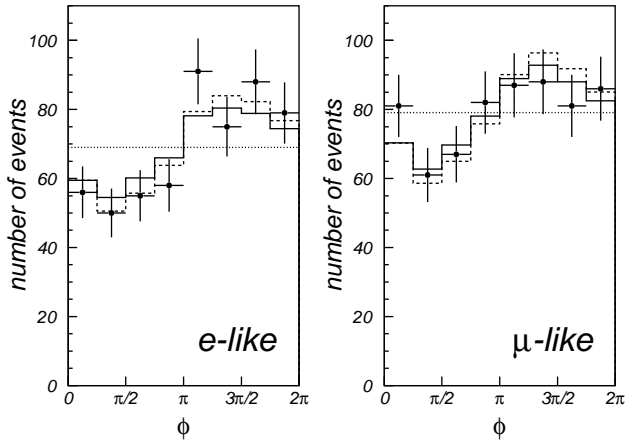}
    \caption{Azimuthal angle distributions of $e$-like and $\mu$-like
      events.  The crosses represent the data points, the histogram
      drawn by solid line (dashed-line) shows the prediction of the
      Monte Carlo based on the flux of [two different flux models]).
      Data are shown with statistical errors. The Monte Carlo has 10
      times more statistic than data. The Monte Carlo histogram is
      normalized to the total number of the real data.  $\phi$
      represents the azimuthal angle. $\phi$ = 0, $\pi/2$, $\pi$ and
      $3\pi/2$ shows particles going to north, west, south, and east,
      respectively. Figure and caption from
      Ref.~\cite{futagami:1999wz}. }
    \label{fig:east-west}
  \end{center}
\end{figure}

In Ref.~\cite{futagami:1999wz} the east-west effect was seen in
neutrinos for the first time.  This effect, first since in cosmic ray
showers, is a direct confirmation of the mostly positively charged
nature of the cosmic rays.  Low-energy cosmic rays are deflected by
the Earth's magnetic field distorting the measured azimuthal spectra.
By selecting a subset of events sensitive to the Earth's magnetic
field this effect was observed as shown in Fig.~\ref{fig:east-west}
taken from \cite{futagami:1999wz}.

While the previous analyses explored only zenith angle distortions
caused by oscillations, by confirming the expected azimuthal
distortion it was demonstrated that the geomagnetic effects on the
production of GeV energy neutrinos was well understood. \\

\noindent
{\bf Tests for non-standard oscillation models~\cite{fukuda:2000np}}

As more data was collected and the techniques were improved much effort
was put into trying to test if any other explanation was as good as
the $\nu_\mu \leftrightarrow \nu_\tau$ at explaining the observed
data.

In Ref.~\cite{fukuda:2000np} with over 1000 days of data the fully
and partially contained events were fit together along with the
upward-going muons and a multi-ring sample which had been enriched
with neutral-current interactions.  In $\nu_\mu \leftrightarrow
\nu_\tau$ oscillations, a simple disappearance of the charged current
events would be expected since tau neutrinos still have neutral
current interactions.  On the other-hand, if the oscillations were to
sterile neutrinos which had no interactions at all, two additional
effects would arise.  First, there would also be a reduction in the
neutral current sample. Second, because of a difference in the forward
scattering cross-sections in the two cases, the normal charge current
oscillations would be suppressed at high energies. This suppression is
known as the ``matter effect.'' A fit was done to the data samples
looking for these effects and was not seen.  Instead, the data fit the
standard $\nu_\mu \leftrightarrow \nu_\tau$ hypothesis extremely well
and excluded oscillations into sterile neutrinos at greater than the
99\% hypothesis level. \\

\noindent
{\bf Observation of the oscillation pattern~\cite{ashie:2004mr}}

In 2004 in Ref.~\cite{ashie:2004mr} the standard oscillation
hypothesis was given strong confirmation by the observation of a
sinusoidal pattern in the oscillation pattern.

\begin{figure}[!htb]
  \begin{center}
    \includegraphics[width=2.90in]{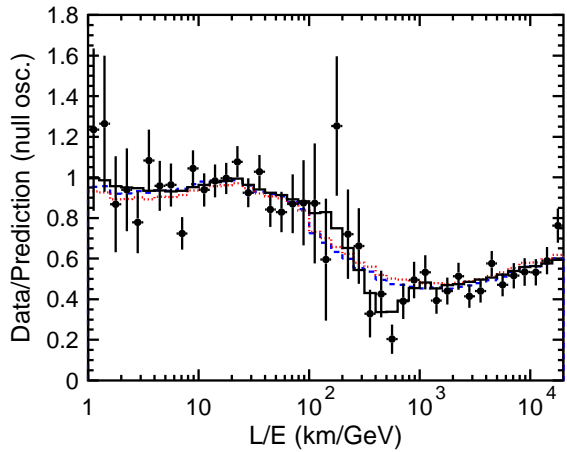}
    \caption{Ratio of the data to the MC events without neutrino
      oscillation (points) as a function of the reconstructed $L/E$
      together with the best-fit expectation for 2-flavor $\nu_\mu
      \leftrightarrow \nu_\tau$ oscillations (solid line).  The error
      bars are statistical only.  Also shown are the best-fit
      expectation for neutrino decay (dashed line) and neutrino
      decoherence (dotted line). Figure and caption from
      Ref.~\cite{ashie:2004mr} }
    \label{fig:bestfit_models}
  \end{center}
\end{figure} 

Since the oscillation equation contains a $\sin$ function, in
principle one should be able to see a sinusoidal dip if the data is
plotted as a function of L/E.  However, in the standard analyses, low
resolution in reconstructed L (distance from production) and E
(neutrino energy) washes out the effect if the data is plotted in
these combinations of variables.  In Ref.~\cite{ashie:2004mr} a
subset of events were selected which had high resolution in these
variables and after plotting the observed suppression relative to the
expectation as a function of L/E the tell-tale dip from oscillations
was observed.  Additionally, by comparing the $\chi^2$ of the data to
the standard oscillation scenario to that of neutrino decay and
decoherence those other exotic hypothesis could be rejected with high
significance.  Figure~\ref{fig:bestfit_models} demonstrating the effect is
taken from Ref.~\cite{ashie:2004mr}. \\

\noindent
{\bf Measurement of atmospheric neutrino parameters~\cite{ashie:2005ik} } \\

In 2005 all of the data from SK-I was refitted. This time the fit was
done jointly with all of the data including upward-going and multiple
ring-samples together.  Ref.~\cite{ashie:2005ik} describes this
analysis in detail along with details of the atmospheric neutrino
Monte Carlo and reconstruction algorithms and reduction.  

%Combined 
\begin{figure}[!htb]
  \begin{center}
    \includegraphics[width=2.9in]{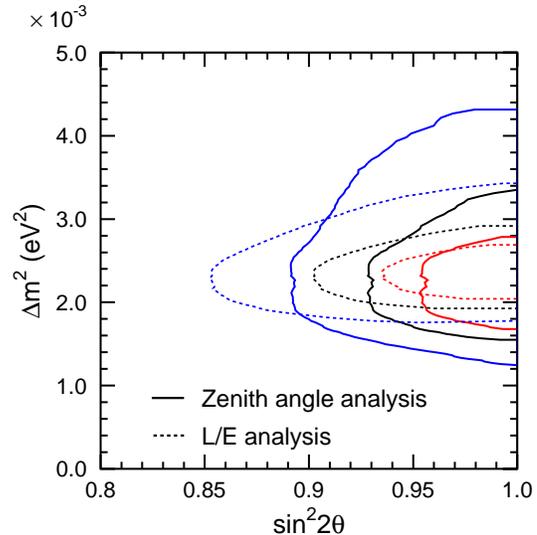}
    \caption{ The 68, 90 and 99\,\% confidence level allowed
      oscillation parameter regions obtained by an $L/E$
      analysis~\cite{ashie:2004mr} and by the 2005 complete data-set
      analysis are compared.  Note the use of the linear y-scale in
      the figure. Figure from Ref~\cite{ashie:2005ik}.}
     \label{fig:allowed-regions-different-int}
  \end{center}
\end{figure}

\begin{figure*}[!ht]
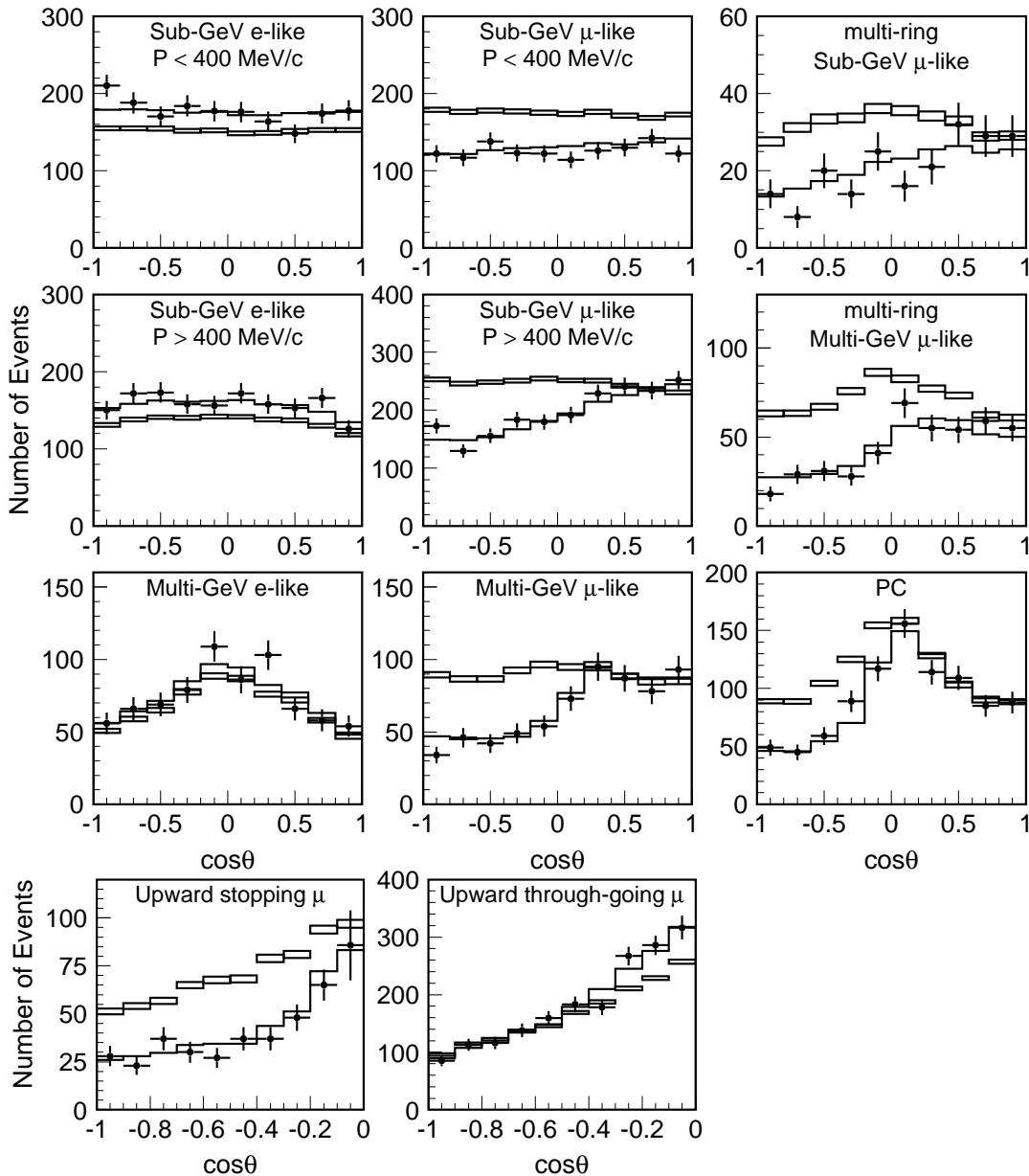

 \begin{center}
    \includegraphics[height=4.7in]{zenith_1a-combined.epsi}
    \includegraphics[height=4.7in]{zenith_2a-combined.epsi} \\
    \hspace{-1.8in}
    \includegraphics[height=1.65in]{zenith_3a-combined.epsi}
    \caption{The zenith angle distribution for fully-contained 1-ring
      events, multi-ring events, partially-contained events and upward
      muons.  The points show the data, box histograms show the
      non-oscillated Monte Carlo events and the lines show the
      best-fit expectations for $\nu_\mu \leftrightarrow \nu_\tau$
      oscillations with $\sin^2 2 \theta = 1.00$ and $\Delta m^2 =
      2.1\times 10^{-3}$~eV$^2$.  The best-fit expectation is
      corrected by the 39 systematic error terms, while the correction
      is not made for the non-oscillated Monte Carlo events.  The
      height of the boxes shows the statistical error of the Monte
      Carlo. Figure and caption from Ref.~\cite{ashie:2005ik}.}
    \label{fig:zenith-combined-plot}
  \end{center}
\end{figure*}
%\clearpage

Thirty-nine systematic uncertainties were accounted for in the fit,
and the best fit for $\nu_\mu \leftrightarrow \nu_\tau$ oscillations
with $\sin^2 2 \theta = 1.00$ and $\Delta m^2 = 2.1\times
10^{-3}$~eV$^2$ was found.  Figure~\ref{fig:zenith-combined-plot},
taken from Ref.~\cite{ashie:2005ik} and shown on the next page,
displays all of the data samples and the results of the fit.  Over
15,000 atmospheric neutrino events were used in the analysis. In
Fig.~\ref{fig:allowed-regions-different-int} the final allowed region
is shown along with the allowed region that is found by the L/E
analysis from Ref.~\cite{fukuda:2000np}.  Both of these analyses give
consistent results.  The better constraint in \sstt in the zenith
angle analysis is due to higher statistics in the sample, while the
better constraint in \dms in the L/E analysis is due to finer binning
in the fit.  Later, as yet unpublished analyses combine the best
feature of both techniques to extract the maximum
information. \\

%Side-by-side option

% \begin{figure*}[!htb]
%   \begin{minipage}[t]{3.0in}
%     \includegraphics[width=2.9in]{contours_comp2.eps}
%     \caption{ The 68, 90 and 99\,\% confidence level allowed
%       oscillation parameter regions obtained by an $L/E$
%       analysis~\cite{ashie:2004mr} and by the combined analysis are
%       compared.  Note the use of the linear y-scale in the
%       figure. Figure from Ref~\cite{ashie:2005ik}.}
%     \label{fig:allowed-regions-different-int}
%   \end{minipage}
%   \hfill
%   \begin{minipage}[t]{3.0in}
%     \includegraphics[width=2.9in]{3flavor-limit-normal.eps}
%     \caption{ Allowed regions of \tonethree as found in the SK-I data
%       set in the three flavor oscillation analysis.  90~\% (thick
%       line) and 99~\% (thin line) confidence level allowed regions are
%       shown in $\Delta m^2$ vs $\sin^2\theta_{13}$.  Normal mass
%       hierarchy ($\Delta m^2$$>$0) is assumed.  The shaded area in the
%       bottom figure shows the region excluded by the CHOOZ reactor
%       neutrino experiment. From Ref~\cite{Hosaka:2006zd}.}
%     \label{fig:3flavor-allowed}
%   \end{minipage}
% \end{figure*}

\noindent
{\bf Three flavor oscillations~\cite{Hosaka:2006zd}} \\

In Ref.~\cite{ashie:2005ik} the data was fit assuming that there were
only oscillations between two flavors of neutrinos.  However, we know
there are three flavors of neutrinos and one can write down a
three-flavor mixing matrix with three mixing angles.  The two that
have been measured (\tonetwo and \ttwothree) control solar and
atmospheric mixing respectively. The third, as yet unmeasured, mixing
angle \tonethree would allow all three of these neutrinos to
oscillate into each-other, producing small amounts of high energy
electron neutrino appearance in the upward going data. The question of
the value of \tonethree is particularly pressing since if it is
extremely small or zero it will not be possible to measure CP
violation in the neutrino sector using the oscillation technique.

\begin{figure}[!htb]
%  \begin{center}
    \hspace{-0.5in}
    \includegraphics[width=2.7in]{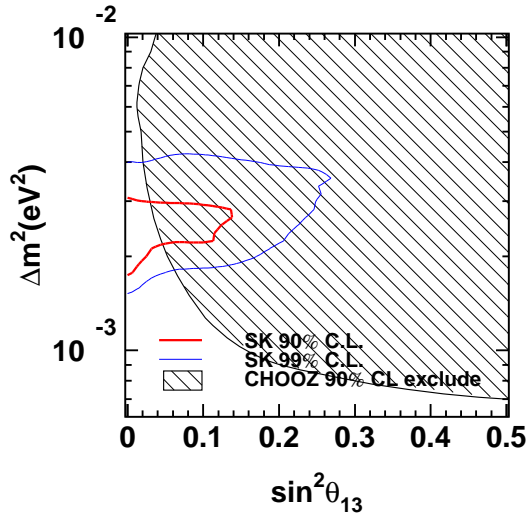}
    \caption{ Allowed regions of \tonethree as found in the SK-I data
      set in the three flavor oscillation analysis.  90~\% (thick
      line) and 99~\% (thin line) confidence level allowed regions are
      shown in $\Delta m^2$ vs $\sin^2\theta_{13}$.  Normal mass
      hierarchy ($\Delta m^2$$>$0) is assumed.  The shaded area in the
      bottom figure shows the region excluded by the CHOOZ reactor
      neutrino experiment. From Ref~\cite{Hosaka:2006zd}.}
    \label{fig:3flavor-allowed}
% \end{center}
\end{figure}

In order to measure or constrain the value of \tonethree using the \SK
data set, a fit was performed using full three flavor oscillation
probabilities, taking into account the resonance effects that can occur
in the earth for such oscillations.  The data was binned so as to be
maximally sensitive to upward going electrons in the relevant energy
regions.  As detailed in Ref.~\cite{Hosaka:2006zd} no evidence for
non-zero \tonethree was found.  A plot showing the \SK \tonethree
allowed region overlaid with the exclusion region for the CHOOZ
reactor experiment is shown in Fig.~\ref{fig:3flavor-allowed}. \\

% Three-flavor

\noindent
{\bf Search for tau lepton appearance~\cite{Abe:2006fu}} \\

None of the previous analyses explicitly searched for the appearance
of the tau lepton from oscillations.  The number of tau leptons in the
data sample is expected to be small.  Only on the order of 80~events
with tau leptons were expected to be produced in fiducial volume of
\superk during the SK-I running period.  In Ref.~\cite{Abe:2006fu}
an explicit search for tau lepton appearance was undertaken.

\begin{figure}[!htb]
  \begin{center}
    \begin{minipage}[b]{2.7in}
      \includegraphics[width=2.0in]{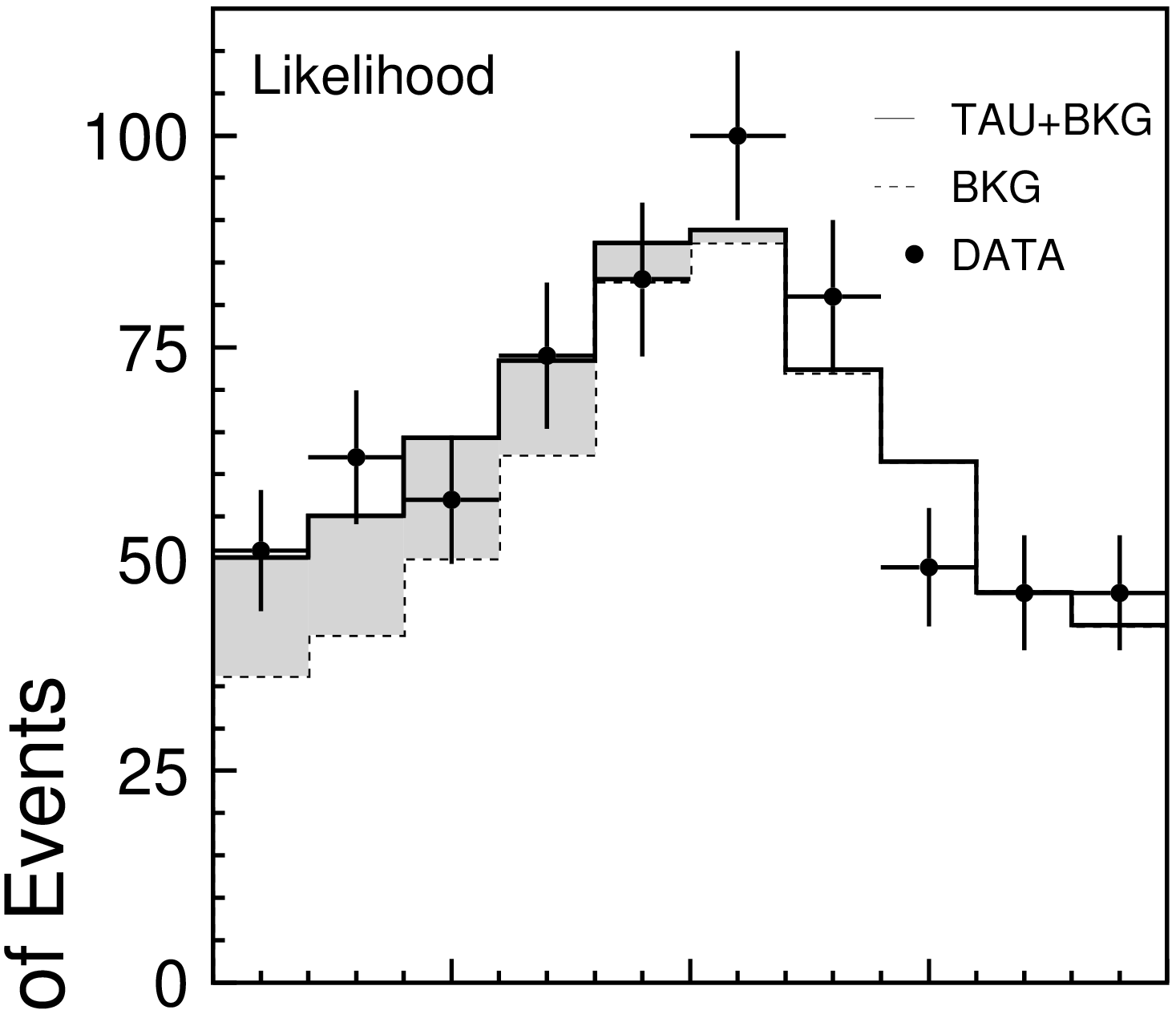}\\
      \includegraphics[width=2.0in]{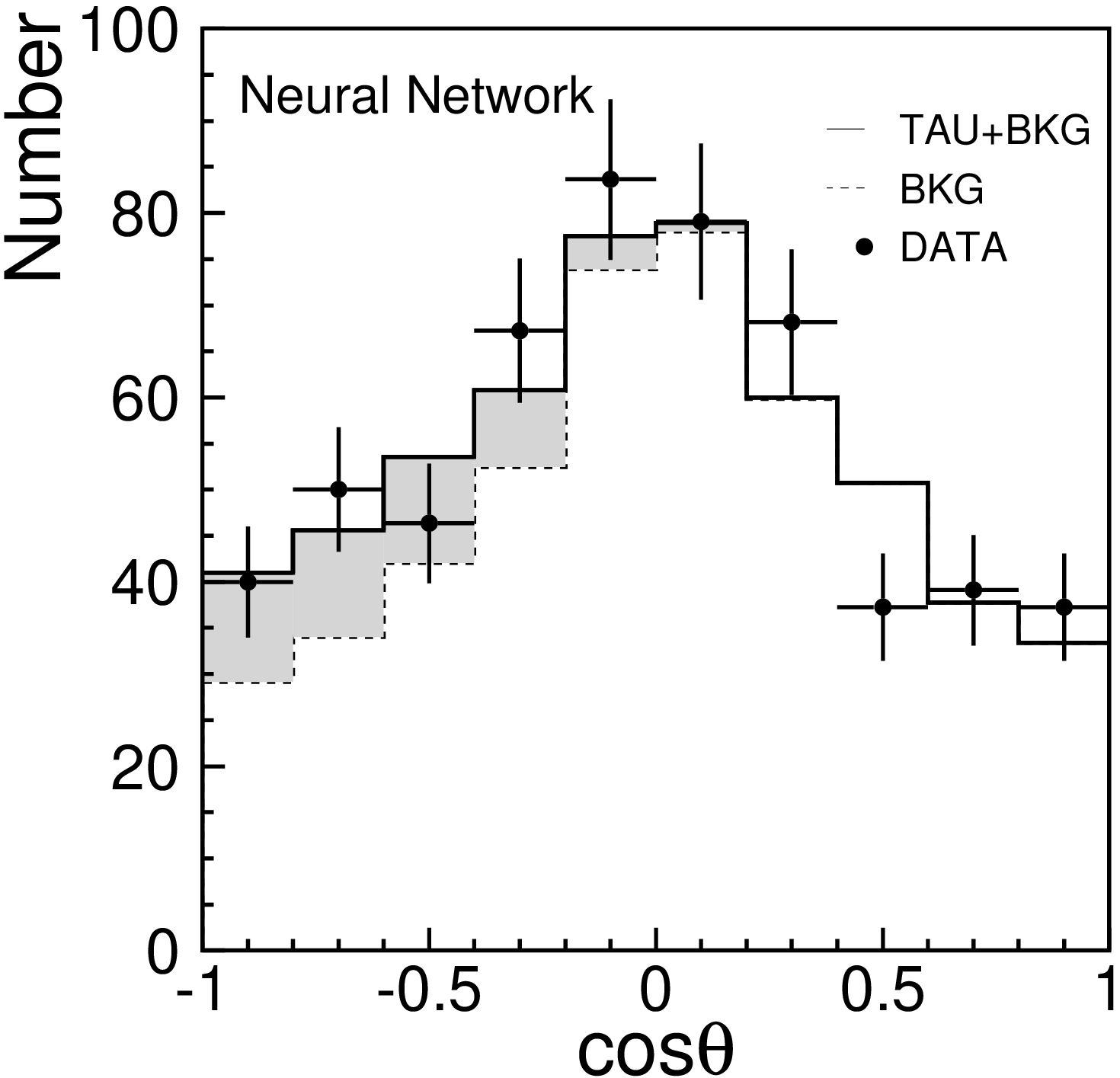}
    \end{minipage} \\
%    \hfill
%    \hspace{.5in}
    \begin{minipage}[b]{2.7in}
      \caption{The zenith angle distributions for the likelihood (top)
        and neural network (bottom) analyses. Zenith angle
        cos$\theta$ = $-$1 (cos$\theta$ = $+$1) indicates upward-going
        (downward-going).  The data sample is fitted after tau
        neutrino event selection criteria are applied. The solid
        histogram shows the best fit including \nutau, and the dashed
        histogram shows the backgrounds from atmospheric neutrinos
        (\nue and \numu).  An excess of tau-like events is observed in
        upward-going direction (shaded area). From
        Ref.~\cite{Abe:2006fu} \label{fig:tau_zenith_fit} }
%      \vspace{1.05in}
    \end{minipage}
  \end{center}
\end{figure}

This search relied on the fact that events containing heavy tau
leptons which decay hadronically, decay symmetrically with more pions
in the final state then the charged current deep-inelastic events
which form the background to the search.  A statistical separation of
signal to background was made using both a likelihood and a
neural-network.  The background was normalized with downward going
events by taking advantage of the fact that the entire tau signal
comes from below.  Fig~\ref{fig:tau_zenith_fit} from
Ref.~\cite{Abe:2006fu} shows the results of the fit.  This analysis
measured a tau appearance signal which was consistent with that
expected from $\nu_\mu
\leftrightarrow \nu_\tau$ oscillations. \\

%\begin{verbatim}
% ATMOS [11]
% ------
% tau       "{A measurement of atmospheric neutrino flux consistent with
% 3-flavor  "{Three flavor neutrino oscillation analysis of atmospheric
% combined  "{A measurement of atmospheric neutrino oscillation
% L/E       "{Evidence for an oscillatory signature in atmospheric
% sterile   "{Tau neutrinos favored over sterile neutrinos in
% stopping  "{Neutrino-induced upward stopping muons in Super-
% east-west "{Observation of the east-west anisotropy of the atmospheric
% upmu      "{Measurement of the flux and zenith-angle distribution of
% evidence  "{Evidence for oscillation of atmospheric neutrinos}",
% mgev      "{Study of the atmospheric neutrino flux in the multi-GeV
% R         "{Measurement of a small atmospheric nu/mu / nu/e ratio}",
%\end{verbatim}

\subsection{Solar Neutrino Oscillations}

The neutrinos produced in the nuclear burning of the sun are of lower
energy than atmospheric neutrinos.  \superk is sensitive mostly to
neutrinos from the ${}^8B$ branch of the pp nuclear fusion chain in
solar burning.  At very low energies the experiment is dominated by
radioactive backgrounds.  However, above approximately 4~MeV the
detector can pick-out the scattering of solar neutrinos off atomic
electrons which make Cherenkov light in the tank.  The ${}^8B$ and
rarer HEP neutrinos have a spectrum which ends near 20~MeV.

\begin{figure}[!htb]
  \begin{center}
    \includegraphics[width=8.6cm]{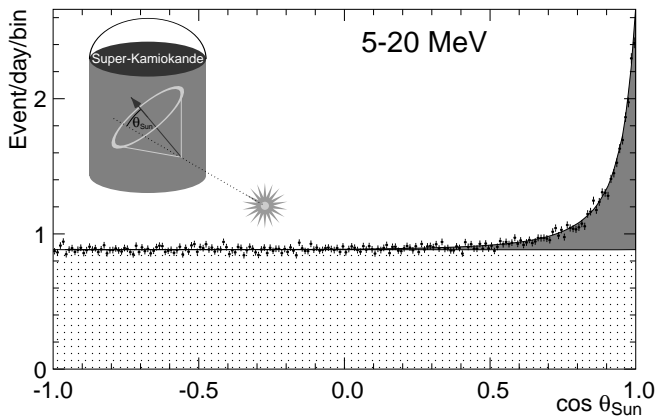}
    \caption{Angular distribution of solar neutrino event candidates.
      The shaded area indicates the elastic scattering peak. The
      dotted area is the contribution from background events.  From
      Ref.~\cite{Hosaka:2005um}}
    \label{fig:cossun}
  \end{center}
\end{figure}

Unlike previous radio-chemical experiments which relied on extraction
of isotopes, \superk collects its data in real time and the electrons
which are scattered by neutrinos point in a direction that is
correlated with the sun.  Therefore, by plotting the direction between
low energy events in the tank and the Sun, one can pick out a peak
containing solar neutrinos.  This is clearly demonstrated in
Fig.~\ref{fig:cossun} which the $\cos$ of the angle between the events
and the sun are plotted.

The first solar neutrino results from \superk were presented in 1998
in Ref.~\cite{fukuda:1998fd} and reported the measured flux from
the first 300 days of data.  In this paper, only the measured rate was
reported and \superk confirmed the existence of deficit in neutrinos
expected from by the sun.  It was consistent with the previous
Kamiokande measurement and reported a flux which was 36\% of the
standard solar model of 1995.  For this first analysis, the energy
threshold was set at 6.5~MeV which was later lowered.

The next two papers used 504 days of data and searched for distinctive
signs of neutrino oscillation.  In Ref~\cite{fukuda:1998rq}, a
measurement was made of the solar flux as measured at \superk in the
day and the night separately.  During the day, neutrinos from the sun
travel down from above, traveling through very little earth.  While
during the night they travel from the other side of the Earth, in
some parts of the year actually passing through the outer core. This
is relevant because, for some regions of oscillation space, a
regeneration of the original flux can happen inside high density
materials. In this paper, no statistically significant sign of this
was seen, thereby excluding parts of the oscillation space previously
allowed by other experiments.  At the same time, an analysis of the
measured energy spectrum shape was published in
Ref.~\cite{fukuda:1998ua} using the 504 day data set.  It was found
that a $\chi^2$ analysis gave a a probability of 4.6\% of being in
agreement with the standard solar model.

In 2001, with more than twice as much data, the analysis threshold was
lowered to 5~MeV and the systematics of the measurement were reduced
due to extensive calibrations and improvements in the analysis
techniques.  In Ref.~\cite{fukuda:2001nk} a precise measurement of
the solar flux was found. Additionally, the spectrum was found to have
no significant energy spectrum distortions, and a no statistically
significant day-night effect was seen. The small but expected seasonal
variation of the flux due to the eccentricity of the Earth's orbit was
observed and new stringent limits on the flux of HEP neutrinos were
presented.  The lack of spectral and zenith angle distortions placed
strong constraints on the oscillation solutions to the solar neutrino
problem and in Ref.~\cite{fukuda:2001nj} an oscillation analysis
was performed using both the spectral and flux information from
\superk.  The result of this analysis was a preference for the Large
Mixing Angle solution.

In 2002 an analysis including all 1496 days of SK-I was
published~\cite{fukuda:2002pe}. The lack of spectral distortion and
daily variation in the flux strongly constrained the regions allowed
for neutrino oscillation from other experiments, leaving only the
high-mass LMA and quasi-vacuum region allowed.  If the \SK interaction
rate and either the standard solar model prediction of the ${}^8B$
flux, or the rates as measured by SNO were included in the analysis, the
LMA angle was uniquely allowed at high confidence level.  This single
allowed region could uniquely explain the solar neutrino problem.

The full SK-I data set was also used to search for other exotic
signals from the Sun. First was the search for anti-electron neutrinos
in Ref.~\cite{gando:2002ub}.  The motivation for this search was to
exclude the possibility that the electron-neutrinos from the sun are
disappearing do to a spin-flavor-precession where neutrinos with a
large magnetic moment are transformed into their anti-particles in the
strong magnetic field of the sun.  Unlike the standard solar neutrino
analysis, the reaction here is inverse beta-decay off of protons in the
nucleus.  No signal was found and a limit was set of 0.8\% of the
standard solar neutrino flux between 8-20~MeV.

Next, in Ref.~\cite{yoo:2003rc}, searches for periodicity in the solar
neutrino data using the Lomb test were performed.  No significant
statistical fluctuation was found.  Finally, in 2004, a search for the
magnetic moment of the neutrino was presented in
Ref.~\cite{liu:2004ny}.  The SK-I data set was used to look for energy
distortion of electron recoil spectrum.  Using \superk's allowed
oscillation parameters a limit of $< 3.6 \times 10^{-10} \mu_B$ was
set. If the constraints on oscillation parameters from other
experiments were also included then the limit was reduced to $ < 1.1
\times 10^{-10} \mu_B$.

\begin{figure}[!htb]
  \begin{center}
    \includegraphics[width=3.0in]{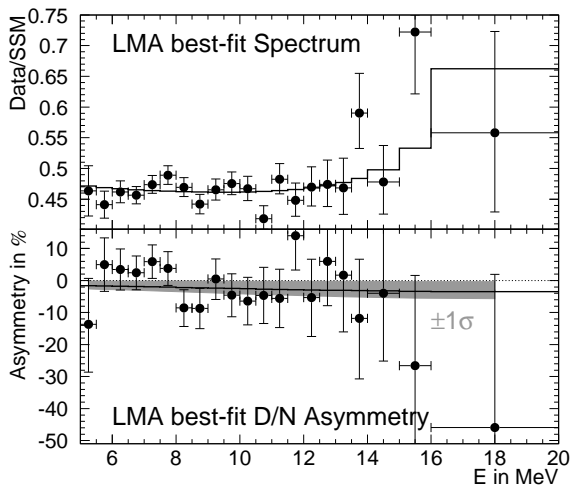}
    \caption{LMA Spectrum (top) and D/N Asymmetry (bottom).  The
      predictions (solid lines) are for $\tan^2\theta=0.55$ and
      $\Delta m^2=6.3\times10^{-5}$eV$^2$ with
      $\phi_{^8B}=0.96\times$Standard Solar Model~[BP2000] and
      $\phi_{\mbox{\tiny\em hep}}=3.6\times$Standard Solar Model.
      Each energy bin is fit independently to the rate (top) and the
      day/night asymmetry (bottom).  The gray bands are the
      $\pm1\sigma$ ranges corresponding to the fitted value over the
      entire range 5-20 MeV: $A=-1.8\pm1.6$\%. Figure and caption from
      Ref.~\cite{smy:2003jf}. }
    \label{fig:lmadnfit}
    \end{center}
\end{figure}

In 2004, in Ref.~\cite{smy:2003jf}, the full SK-I data set was once
again analyzed but this time with a maximum likelihood analysis
applied to the zenith angle dependence of the data.  This lowered the
statistical uncertainty on the day/night ratio by 25\% compared to the
previous measurement.  This was equivalent to adding three years of
live-time running if still using the previous method.
Fig.~\ref{fig:lmadnfit} from Ref.~\cite{smy:2003jf} shows the data
and best fit spectrum in the LMA region along with the D/N asymmetry
as a function of energy.

In 2005 a full detailed description of the SK-I solar neutrino
analysis was published in Ref.~\cite{Hosaka:2005um}. This paper
included descriptions of the simulation, reconstruction and analysis
techniques.  In the large mixing angle region the day-night asymmetry
was found to be:
\begin{align}
A & = { (\Phi_D-\Phi_N) \over {1 \over 2}(\Phi_D+\Phi_N)} =  \notag\\
  & -1.7\% \pm 1.6\% (stat.)^{+1.3\%}_{-1.2\%} (sys.) \pm .04\% (\Delta m^2)
\end{align}

\noindent
which is statistically compatible with zero and is to be compared with
the expected asymmetry which ranges from -1.7\% to 1.0\%.

%   \begin{center}
%     \includegraphics[width=3.5in]{spect.eps}
%     \caption{Energy spectrum of the solar neutrino signal.  The
%       horizontal axis is the total energy of the recoil electrons.
%       The vertical axis is the event rate of the observed solar
%       neutrino signal events.  The error bars are a quadrature of the
%       statistical and uncorrelated errors.  The reduction efficiencies
%       in Fig.~[Another Figure] are corrected.  BP2004 1SSM flux
%       values are used for the $^8$B and hep MC fluxes in this plot.
%       The dashed line shows the contribution of only $^8$B. Figure and
%       caption from Ref.~\cite{Hosaka:2005um}. }
%     \label{fig:spec}
%   \end{center}
% \end{figure}

% \begin{verbatim}
% SOLAR [11]
% ------
% SKI         "{Solar neutrino measurements in Super-Kamiokande-I}",
% mag-mom     "{Limit on the neutrino magnetic moment using 1496 days of
% day-nightII "{Precise measurement of the solar neutrino day/night and
% modulation  "{A search for periodic modulations of the solar neutrino
% anti-nue    "{Search for anti-nu/e from the sun at Super-Kamiokande-I}",
% 1496        "{Determination of solar neutrino oscillation parameters
% OSC analy   "{Solar B-8 and he p neutrino measurements from 1258 days of
% 1298 measure"{Constraints on neutrino oscillations using 1258 days of
% day-night   "{Constraints on neutrino oscillation parameters from the
% spectrum    "{Measurement of the solar neutrino energy spectrum using
% flux-300    "{Measurements of the solar neutrino flux from Super-
% \end{verbatim}

\subsection{The Search for Proton Decay}

\SK has also set important limits on the decay of the nucleon.  The
limits from \SK have ruled out SU(5) and the minimal
super-symmetric model.  The published limits on 
$p \rightarrow e^+ \pi^0$ as predicted in SU(5) are found in
Ref.~\cite{shiozawa:1998si}. 

Many models of super-symmetry predict that the nucleon should decay
with strange quarks in the final state.  The first 535 days of SK data
were analyzed to search for \nukp and the limits were presented in
Ref.~\cite{hayato:1999az}.  In 2005 the full SK-I data set was
employed to search for SUSY mediated decays of the proton.  In the
analysis presented in Ref.~\cite{kobayashi:2005pe} stringent limits
were set on \nukp, \nukz, \mukz and \ekz modes.

A related search for SUSY generated physics is the search for neutral
Q-Balls as presented in Ref.~\cite{Takenaga:2006nr}.  Q-balls are
topological solitons predicted in some SUSY models and interact as
they pass through the \SK tank leaving a trail of pions in their wake.
A search was carried out using 542 days of the SK-II data set. No
evidence for Q-balls were found, and \SK finds the lowest limits in the
world for Q-ball cross-sections below 200~mb.

% \begin{verbatim}
% PDK [4]
% ----
% qball     "Search for neutral Q-balls in Super-Kamiokande II",
% K+?      "{Search for nucleon decay via modes favored by
% K+       "{Search for proton decay through p $-$$>$ anti-nu K+ in a
% e+pi0    "{Search for proton decay via p --> e+ pi0 in a large water
% \end{verbatim}

\subsection{The Search for Astrophysical Phenomenon}

\superk has been used to search for several sources of neutrinos from
outside of our solar system. Gamma Ray Bursters  are among the
most luminous sources that have ever been observed in the universe.
\SK performed a search for neutrinos that were in coincidence with the
BATSE detector located on NASA's Compton Gamma Ray Observatory.  The
entire \SK data sample from 7~MeV to $\sim$ 100~TeV was compared with the
BATSE online catalog in the period of April 1996 to May of 2000.  The
results were presented in Ref.~\cite{fukuda:2002nf} and no
statistically significant signal was found. 

A supernova which occurred in the center of our galaxy would produce
on the order of 10,000 interactions inside the \superk tank, yielding
a rich sample of events for analysis.  In the period of \SK running
there has been no observed galactic supernova.  In
Ref.~\cite{Ikeda:2007sa} a limit on the rate of core-collapse
supernovae out to 100~kiloparsecs using the SK-I and SK-II data sets
was reported to be $< 0.32/{\rm year}$.  Although no core collapse
was detected by \superk, it is believed that the universe should be
bathed in the relic neutrinos of all of the supernovas that have
exploded in the past.  In Ref.~\cite{malek:2002ns} the SK-I data
sample was examined for such events.  There is a region of energy
between the endpoint of solar the solar neutrino spectrum and before
the decay electrons of cosmic ray muons where one can hope to see the
small number of expected events.  No excess over background was
observed, setting a limit just above the expected signal from many
models.

The upward-going muons used for the atmospheric oscillation analysis
can also be used to search for high energy sources of neutrinos.
The only muons that come from below the tank are generated by
neutrinos, and the neutrinos that produce them range from approximately
100~GeV to 100~TeV.

\begin{figure}[!htb]
  \begin{center}
    \center{\includegraphics[width=2.75in]{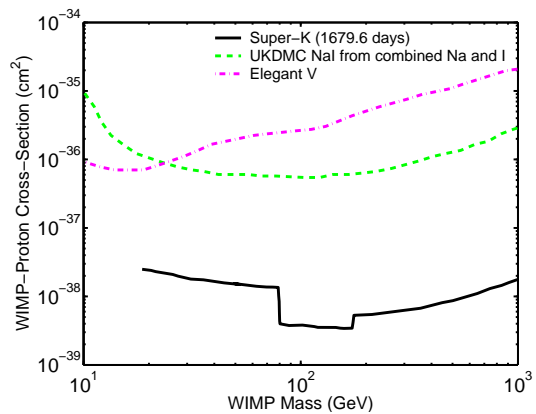}}
    \caption{Super-K 90 \% CL exclusion region in WIMP parameter space
      for a WIMP with spin-dependent coupling along with corresponding
      90 \% CL exclusion limits from UKDMC (dashed) and ELEGANT
      (dot-dashed). Figure and caption from
      Ref.~\cite{desai:2004pq}. }
    \label{fig:wimp_limit}
  \end{center}
\end{figure}

In addition to looking for neutrinos from astrophysical objects there
is an expected neutrino signal from WIMP dark matter.  Since WIMPs
undergo gravitational attraction they are expected to gravitate around
heavy bodies such as the Earth and the Sun.  If a WIMP and anti-WIMP
annihilate near the center of one of these objects where they have
accumulated, they will decay into standard model particles some of which
decay into neutrinos.

Therefore, if there is an excess signal of upward going neutrinos
coming from either the center of the Earth or the Sun it can be
interpreted as a signal for dark matter.  It is shown in
Ref.~\cite{desai:2004pq} that the limit on spin-dependent couplings
inferred from the lack of an excess in the direction of the sun is on
the order of 100 times more sensitive than comparable terrestrial
direct dark matter experiments.  Fig.~\ref{fig:wimp_limit} taken from
Ref.~\cite{desai:2004pq} shows a comparison with the \SK limits
with that of other direct-detection experiments.

Additionally, several pure astronomical analyses have been performed
with \SK using samples of both downward-going and upward-going muons.
In Ref.~\cite{Guillian:2005wp} the anisotropy of the primary cosmic
ray flux with a mean energy of 10~TeV was measured by studying the
relative sidereal variation of downward-going muons in
the \SK tank.

By using upward-going muons, searches were performed for sources of
high energy astrophysical neutrinos.  In Ref.~\cite{Abe:2006at} the
sample of upward-going muons were used to search the sky for point
sources and signatures of a diffuse flux of neutrinos from the
galactic plane.  Additional time correlated searches with some known
sources were also performed.  Above 1~TeV of muon energy, the flux of
upward going muons from atmospheric neutrinos becomes small enough to
look for a diffuse flux of extremely high energy neutrinos from
astrophysical sources such as active galactic nuclei.  In
Ref.~\cite{Swanson:2006gm} the outer detector of \SK was used to
measure the direction of extremely high energy events. One event was
found which was consistent with the background expectation and limits
were set on this flux.

% \begin{verbatim}
% ASTRO [6]
% ------
% ultra-he   "Search for diffuse astrophysical neutrino flux using ultra-
% muon-astro "{High energy neutrino astronomy using upward-going muons in
% Anisot     "Observation of the anisotropy of 10-TeV primary cosmic ray
% wimps      "{Search for dark matter WIMPs using upward through-going
% relic      "{Search for supernova relic neutrinos at Super-Kamiokande}",
% GRB        "{Search for neutrinos from gamma-ray bursts using Super-

% \end{verbatim}

\section{Conclusions}

This review of the published work of the \superk collaboration has
shown the depth and importance of the work that the experiment has
achieved.  From astronomy and astrophysics, to tests of grand unified
models, and to the observation of neutrino mass using neutrinos from
the atmosphere and the Sun, \SK has has a major impact on particle
physics.  

It would be a mistake to take this snapshot of the published work of
\superk as its final word.  There is a large data set from
SK-II which is currently being analyzed, and will soon be published.
These analyses will increase the precision of the parameters \superk has
already measured. In addition, the larger data set will allow analyses
for more subtle effects which have not yet been studied.  

\begin{figure}[!htb]
  \centering
  \includegraphics[width=2.50in]{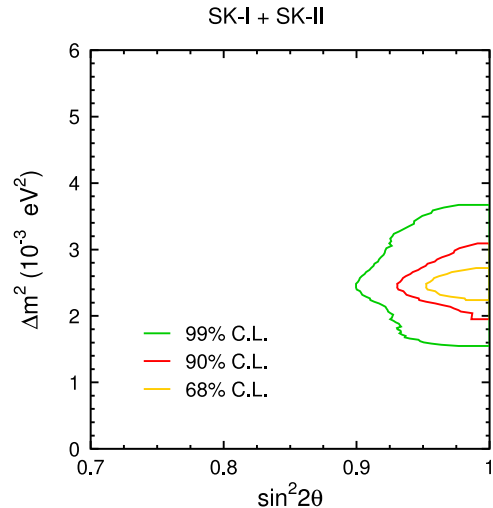}
  \caption{Preliminary allowed region  of atmospheric oscillation parameters
    using the combined data sets of SK-I and SK-II.  The fit to the data
    employed 380 bins, 70 systematic errors and had a Chi-squared
    probability of 18\%.  Note the use of the linear y-scale in the figure.}
  \label{fig:skI+II-region}
\end{figure}

As one example of this, Fig.~\ref{fig:skI+II-region} shows the latest
result of the atmospheric oscillation analysis using the combined SK-I
and SK-II data sets.  In addition to the increase in statistics this
analysis has improvements in the binning and systematic errors.  It
should be noted that in this figure the y-axis is plotted on a linear
scale.

This analysis employed 380 bins and 70 systematic errors.  The best
fit to the atmospheric oscillation parameters were:
\begin{equation}
  \Delta m^2 = 2.5\times 10^{-3}~eV^2 \quad  \sin^2 2 \theta = 1.00,
\end{equation}
with the allowed region at 90\% CL being equal to:
\begin{equation}
   1.9\times 10^{-3}~eV^2 < \Delta m^2 < 3.1\times 10^{-3}~eV^2 \quad
   \sin^2 2 \theta < 0.93.
\end{equation}
% \begin{verbatim}

\begin{figure*}[!htb]
  \centering
  \includegraphics[width=3.0in,angle=-90]{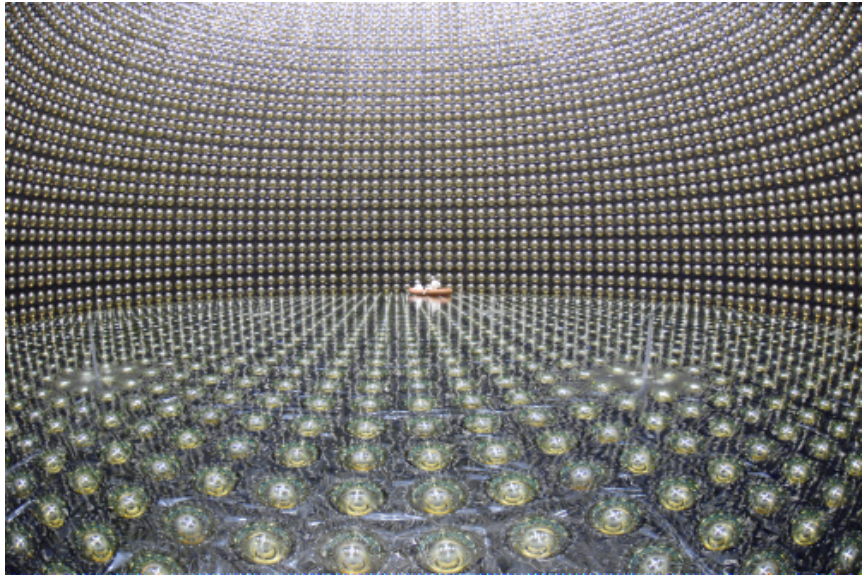}
  \caption{Picture of SK-III during filling and before the beginning
    of operations. {\it Photo-credit: Kamioka Observatory,
      ICRR(Institute for Cosmic Ray Research), The University of
      Tokyo.} }
  \label{fig:SKIII_figure}
\end{figure*}

Since June of 2006 SK-III has been operational and taking data.
Fig~\ref{fig:SKIII_figure} is a picture of SK-III during filling
before operations commenced.  \superk will continue to collect data,
wait for supernovae, and in 2009 will once again become a target for
an accelerator based neutrino beam.

\section*{Acknowledgments}

The author and the \superk collaboration gratefully acknowledge the
cooperation of the Kamioka Mining and Smelting Company.  The
Super-Kamiokande experiment has been built and operated from funding
by the Japanese Ministry of Education, Culture, Sports, Science and
Technology, the United States Department of Energy, and the U.S.
National Science Foundation.

\bibliography{external-refs,sk-refs}

\begin{thebibliography}{44}
\expandafter\ifx\csname natexlab\endcsname\relax\def\natexlab#1{#1}\fi
\expandafter\ifx\csname bibnamefont\endcsname\relax
  \def\bibnamefont#1{#1}\fi
\expandafter\ifx\csname bibfnamefont\endcsname\relax
  \def\bibfnamefont#1{#1}\fi
\expandafter\ifx\csname citenamefont\endcsname\relax
  \def\citenamefont#1{#1}\fi
\expandafter\ifx\csname url\endcsname\relax
  \def\url#1{\texttt{#1}}\fi
\expandafter\ifx\csname urlprefix\endcsname\relax\def\urlprefix{URL }\fi
\providecommand{\bibinfo}[2]{#2}
\providecommand{\eprint}[2][]{\url{#2}}

\bibitem[{\citenamefont{Hirata et~al.}(1988)}]{Hirata:1988uy}
\bibinfo{author}{\bibfnamefont{K.~S.} \bibnamefont{Hirata}}
  \bibnamefont{et~al.} (\bibinfo{collaboration}{KAMIOKANDE-II}),
  \bibinfo{journal}{Phys. Lett.} \textbf{\bibinfo{volume}{B205}},
  \bibinfo{pages}{416} (\bibinfo{year}{1988}).

\bibitem[{\citenamefont{Hirata et~al.}(1992)}]{Hirata:1992ku}
\bibinfo{author}{\bibfnamefont{K.~S.} \bibnamefont{Hirata}}
  \bibnamefont{et~al.} (\bibinfo{collaboration}{Kamiokande-II}),
  \bibinfo{journal}{Phys. Lett.} \textbf{\bibinfo{volume}{B280}},
  \bibinfo{pages}{146} (\bibinfo{year}{1992}).

\bibitem[{\citenamefont{Casper et~al.}(1991)}]{Casper:1991ac}
\bibinfo{author}{\bibfnamefont{D.}~\bibnamefont{Casper}} \bibnamefont{et~al.},
  \bibinfo{journal}{Phys. Rev. Lett.} \textbf{\bibinfo{volume}{66}},
  \bibinfo{pages}{2561} (\bibinfo{year}{1991}).

\bibitem[{\citenamefont{Becker-Szendy et~al.}(1992)}]{Becker-Szendy:1992hq}
\bibinfo{author}{\bibfnamefont{R.}~\bibnamefont{Becker-Szendy}}
  \bibnamefont{et~al.}, \bibinfo{journal}{Phys. Rev.}
  \textbf{\bibinfo{volume}{D46}}, \bibinfo{pages}{3720} (\bibinfo{year}{1992}).

\bibitem[{\citenamefont{Georgi and Glashow}(1974)}]{Georgi:1974sy}
\bibinfo{author}{\bibfnamefont{H.}~\bibnamefont{Georgi}} \bibnamefont{and}
  \bibinfo{author}{\bibfnamefont{S.~L.} \bibnamefont{Glashow}},
  \bibinfo{journal}{Phys. Rev. Lett.} \textbf{\bibinfo{volume}{32}},
  \bibinfo{pages}{438} (\bibinfo{year}{1974}).

\bibitem[{\citenamefont{Aglietta et~al.}(1989)}]{Aglietta:1988be}
\bibinfo{author}{\bibfnamefont{M.}~\bibnamefont{Aglietta}} \bibnamefont{et~al.}
  (\bibinfo{collaboration}{The NUSEX}), \bibinfo{journal}{Europhys. Lett.}
  \textbf{\bibinfo{volume}{8}}, \bibinfo{pages}{611} (\bibinfo{year}{1989}).

\bibitem[{\citenamefont{Daum et~al.}(1995)}]{Daum:1994bf}
\bibinfo{author}{\bibfnamefont{K.}~\bibnamefont{Daum}} \bibnamefont{et~al.}
  (\bibinfo{collaboration}{Frejus.}), \bibinfo{journal}{Z. Phys.}
  \textbf{\bibinfo{volume}{C66}}, \bibinfo{pages}{417} (\bibinfo{year}{1995}).

\bibitem[{\citenamefont{Fukuda et~al.}(2003)}]{fukuda:2002uc}
\bibinfo{author}{\bibfnamefont{Y.}~\bibnamefont{Fukuda}} \bibnamefont{et~al.},
  \bibinfo{journal}{Nucl. Instrum. Meth.} \textbf{\bibinfo{volume}{A501}},
  \bibinfo{pages}{418} (\bibinfo{year}{2003}).

\bibitem[{\citenamefont{Nakahata et~al.}(1999)}]{nakahata:1998pz}
\bibinfo{author}{\bibfnamefont{M.}~\bibnamefont{Nakahata}} \bibnamefont{et~al.}
  (\bibinfo{collaboration}{Super-Kamiokande}), \bibinfo{journal}{Nucl. Instrum.
  Meth.} \textbf{\bibinfo{volume}{A421}}, \bibinfo{pages}{113}
  (\bibinfo{year}{1999}), \eprint{hep-ex/9807027}.

\bibitem[{\citenamefont{Blaufuss et~al.}(2001)}]{blaufuss:2000tp}
\bibinfo{author}{\bibfnamefont{E.}~\bibnamefont{Blaufuss}} \bibnamefont{et~al.}
  (\bibinfo{collaboration}{Super-Kamiokande}), \bibinfo{journal}{Nucl. Instrum.
  Meth.} \textbf{\bibinfo{volume}{A458}}, \bibinfo{pages}{638}
  (\bibinfo{year}{2001}), \eprint{hep-ex/0005014}.

\bibitem[{\citenamefont{Takeuchi et~al.}(1999)}]{takeuchi:1999zq}
\bibinfo{author}{\bibfnamefont{Y.}~\bibnamefont{Takeuchi}} \bibnamefont{et~al.}
  (\bibinfo{collaboration}{SuperKamiokade}), \bibinfo{journal}{Phys. Lett.}
  \textbf{\bibinfo{volume}{B452}}, \bibinfo{pages}{418} (\bibinfo{year}{1999}),
  \eprint{hep-ex/9903006}.

\bibitem[{\citenamefont{Ashie et~al.}(2005)}]{ashie:2005ik}
\bibinfo{author}{\bibfnamefont{Y.}~\bibnamefont{Ashie}} \bibnamefont{et~al.}
  (\bibinfo{collaboration}{Super-Kamiokande}), \bibinfo{journal}{Phys. Rev.}
  \textbf{\bibinfo{volume}{D71}}, \bibinfo{pages}{112005}
  (\bibinfo{year}{2005}), \eprint{hep-ex/0501064}.

\bibitem[{\citenamefont{Hosaka et~al.}(2006{\natexlab{a}})}]{Hosaka:2005um}
\bibinfo{author}{\bibfnamefont{J.}~\bibnamefont{Hosaka}} \bibnamefont{et~al.}
  (\bibinfo{collaboration}{Super-Kamkiokande}), \bibinfo{journal}{Phys. Rev.}
  \textbf{\bibinfo{volume}{D73}}, \bibinfo{pages}{112001}
  (\bibinfo{year}{2006}{\natexlab{a}}), \eprint{hep-ex/0508053}.

\bibitem[{\citenamefont{Fukuda et~al.}(1998{\natexlab{a}})}]{fukuda:1998tw}
\bibinfo{author}{\bibfnamefont{Y.}~\bibnamefont{Fukuda}} \bibnamefont{et~al.}
  (\bibinfo{collaboration}{Super-Kamiokande}), \bibinfo{journal}{Phys. Lett.}
  \textbf{\bibinfo{volume}{B433}}, \bibinfo{pages}{9}
  (\bibinfo{year}{1998}{\natexlab{a}}), \eprint{hep-ex/9803006}.

\bibitem[{\citenamefont{Fukuda et~al.}(1998{\natexlab{b}})}]{fukuda:1998ub}
\bibinfo{author}{\bibfnamefont{Y.}~\bibnamefont{Fukuda}} \bibnamefont{et~al.}
  (\bibinfo{collaboration}{Super-Kamiokande}), \bibinfo{journal}{Phys. Lett.}
  \textbf{\bibinfo{volume}{B436}}, \bibinfo{pages}{33}
  (\bibinfo{year}{1998}{\natexlab{b}}), \eprint{hep-ex/9805006}.

\bibitem[{\citenamefont{Fukuda et~al.}(1998{\natexlab{c}})}]{fukuda:1998mi}
\bibinfo{author}{\bibfnamefont{Y.}~\bibnamefont{Fukuda}} \bibnamefont{et~al.}
  (\bibinfo{collaboration}{Super-Kamiokande}), \bibinfo{journal}{Phys. Rev.
  Lett.} \textbf{\bibinfo{volume}{81}}, \bibinfo{pages}{1562}
  (\bibinfo{year}{1998}{\natexlab{c}}), \eprint{hep-ex/9807003}.

\bibitem[{\citenamefont{Fukuda et~al.}(1999{\natexlab{a}})}]{fukuda:1998ah}
\bibinfo{author}{\bibfnamefont{Y.}~\bibnamefont{Fukuda}} \bibnamefont{et~al.}
  (\bibinfo{collaboration}{Super-Kamiokande}), \bibinfo{journal}{Phys. Rev.
  Lett.} \textbf{\bibinfo{volume}{82}}, \bibinfo{pages}{2644}
  (\bibinfo{year}{1999}{\natexlab{a}}), \eprint{hep-ex/9812014}.

\bibitem[{\citenamefont{Fukuda et~al.}(1999{\natexlab{b}})}]{fukuda:1999pp}
\bibinfo{author}{\bibfnamefont{Y.}~\bibnamefont{Fukuda}} \bibnamefont{et~al.}
  (\bibinfo{collaboration}{Super-Kamiokande}), \bibinfo{journal}{Phys. Lett.}
  \textbf{\bibinfo{volume}{B467}}, \bibinfo{pages}{185}
  (\bibinfo{year}{1999}{\natexlab{b}}), \eprint{hep-ex/9908049}.

\bibitem[{\citenamefont{Futagami et~al.}(1999)}]{futagami:1999wz}
\bibinfo{author}{\bibfnamefont{T.}~\bibnamefont{Futagami}} \bibnamefont{et~al.}
  (\bibinfo{collaboration}{Super-Kamiokande}), \bibinfo{journal}{Phys. Rev.
  Lett.} \textbf{\bibinfo{volume}{82}}, \bibinfo{pages}{5194}
  (\bibinfo{year}{1999}), \eprint{astro-ph/9901139}.

\bibitem[{\citenamefont{Fukuda et~al.}(2000)}]{fukuda:2000np}
\bibinfo{author}{\bibfnamefont{S.}~\bibnamefont{Fukuda}} \bibnamefont{et~al.}
  (\bibinfo{collaboration}{Super-Kamiokande}), \bibinfo{journal}{Phys. Rev.
  Lett.} \textbf{\bibinfo{volume}{85}}, \bibinfo{pages}{3999}
  (\bibinfo{year}{2000}), \eprint{hep-ex/0009001}.

\bibitem[{\citenamefont{Ashie et~al.}(2004)}]{ashie:2004mr}
\bibinfo{author}{\bibfnamefont{Y.}~\bibnamefont{Ashie}} \bibnamefont{et~al.}
  (\bibinfo{collaboration}{Super-Kamiokande}), \bibinfo{journal}{Phys. Rev.
  Lett.} \textbf{\bibinfo{volume}{93}}, \bibinfo{pages}{101801}
  (\bibinfo{year}{2004}), \eprint{hep-ex/0404034}.

\bibitem[{\citenamefont{Hosaka et~al.}(2006{\natexlab{b}})}]{Hosaka:2006zd}
\bibinfo{author}{\bibfnamefont{J.}~\bibnamefont{Hosaka}} \bibnamefont{et~al.}
  (\bibinfo{collaboration}{Super-Kamiokande}), \bibinfo{journal}{Phys. Rev.}
  \textbf{\bibinfo{volume}{D74}}, \bibinfo{pages}{032002}
  (\bibinfo{year}{2006}{\natexlab{b}}), \eprint{hep-ex/0604011}.

\bibitem[{\citenamefont{Abe et~al.}(2006{\natexlab{a}})}]{Abe:2006fu}
\bibinfo{author}{\bibfnamefont{K.}~\bibnamefont{Abe}} \bibnamefont{et~al.}
  (\bibinfo{collaboration}{Super-Kamiokande}), \bibinfo{journal}{Phys. Rev.
  Lett.} \textbf{\bibinfo{volume}{97}}, \bibinfo{pages}{171801}
  (\bibinfo{year}{2006}{\natexlab{a}}), \eprint{hep-ex/0607059}.

\bibitem[{\citenamefont{Fukuda et~al.}(1998{\natexlab{d}})}]{fukuda:1998fd}
\bibinfo{author}{\bibfnamefont{Y.}~\bibnamefont{Fukuda}} \bibnamefont{et~al.}
  (\bibinfo{collaboration}{Super-Kamiokande}), \bibinfo{journal}{Phys. Rev.
  Lett.} \textbf{\bibinfo{volume}{81}}, \bibinfo{pages}{1158}
  (\bibinfo{year}{1998}{\natexlab{d}}), \eprint{hep-ex/9805021}.

\bibitem[{\citenamefont{Fukuda et~al.}(1999{\natexlab{c}})}]{fukuda:1998rq}
\bibinfo{author}{\bibfnamefont{Y.}~\bibnamefont{Fukuda}} \bibnamefont{et~al.}
  (\bibinfo{collaboration}{Super-Kamiokande}), \bibinfo{journal}{Phys. Rev.
  Lett.} \textbf{\bibinfo{volume}{82}}, \bibinfo{pages}{1810}
  (\bibinfo{year}{1999}{\natexlab{c}}), \eprint{hep-ex/9812009}.

\bibitem[{\citenamefont{Fukuda et~al.}(1999{\natexlab{d}})}]{fukuda:1998ua}
\bibinfo{author}{\bibfnamefont{Y.}~\bibnamefont{Fukuda}} \bibnamefont{et~al.}
  (\bibinfo{collaboration}{Super-Kamiokande}), \bibinfo{journal}{Phys. Rev.
  Lett.} \textbf{\bibinfo{volume}{82}}, \bibinfo{pages}{2430}
  (\bibinfo{year}{1999}{\natexlab{d}}), \eprint{hep-ex/9812011}.

\bibitem[{\citenamefont{Fukuda et~al.}(2001{\natexlab{a}})}]{fukuda:2001nk}
\bibinfo{author}{\bibfnamefont{S.}~\bibnamefont{Fukuda}} \bibnamefont{et~al.}
  (\bibinfo{collaboration}{Super-Kamiokande}), \bibinfo{journal}{Phys. Rev.
  Lett.} \textbf{\bibinfo{volume}{86}}, \bibinfo{pages}{5656}
  (\bibinfo{year}{2001}{\natexlab{a}}), \eprint{hep-ex/0103033}.

\bibitem[{\citenamefont{Fukuda et~al.}(2001{\natexlab{b}})}]{fukuda:2001nj}
\bibinfo{author}{\bibfnamefont{S.}~\bibnamefont{Fukuda}} \bibnamefont{et~al.}
  (\bibinfo{collaboration}{Super-Kamiokande}), \bibinfo{journal}{Phys. Rev.
  Lett.} \textbf{\bibinfo{volume}{86}}, \bibinfo{pages}{5651}
  (\bibinfo{year}{2001}{\natexlab{b}}), \eprint{hep-ex/0103032}.

\bibitem[{\citenamefont{Fukuda et~al.}(2002{\natexlab{a}})}]{fukuda:2002pe}
\bibinfo{author}{\bibfnamefont{S.}~\bibnamefont{Fukuda}} \bibnamefont{et~al.}
  (\bibinfo{collaboration}{Super-Kamiokande}), \bibinfo{journal}{Phys. Lett.}
  \textbf{\bibinfo{volume}{B539}}, \bibinfo{pages}{179}
  (\bibinfo{year}{2002}{\natexlab{a}}), \eprint{hep-ex/0205075}.

\bibitem[{\citenamefont{Gando et~al.}(2003)}]{gando:2002ub}
\bibinfo{author}{\bibfnamefont{Y.}~\bibnamefont{Gando}} \bibnamefont{et~al.}
  (\bibinfo{collaboration}{Super-Kamiokande}), \bibinfo{journal}{Phys. Rev.
  Lett.} \textbf{\bibinfo{volume}{90}}, \bibinfo{pages}{171302}
  (\bibinfo{year}{2003}), \eprint{hep-ex/0212067}.

\bibitem[{\citenamefont{Yoo et~al.}(2003)}]{yoo:2003rc}
\bibinfo{author}{\bibfnamefont{J.}~\bibnamefont{Yoo}} \bibnamefont{et~al.}
  (\bibinfo{collaboration}{Super-Kamiokande}), \bibinfo{journal}{Phys. Rev.}
  \textbf{\bibinfo{volume}{D68}}, \bibinfo{pages}{092002}
  (\bibinfo{year}{2003}), \eprint{hep-ex/0307070}.

\bibitem[{\citenamefont{Liu et~al.}(2004)}]{liu:2004ny}
\bibinfo{author}{\bibfnamefont{D.~W.} \bibnamefont{Liu}} \bibnamefont{et~al.}
  (\bibinfo{collaboration}{Super-Kamiokande}), \bibinfo{journal}{Phys. Rev.
  Lett.} \textbf{\bibinfo{volume}{93}}, \bibinfo{pages}{021802}
  (\bibinfo{year}{2004}), \eprint{hep-ex/0402015}.

\bibitem[{\citenamefont{Smy et~al.}(2004)}]{smy:2003jf}
\bibinfo{author}{\bibfnamefont{M.~B.} \bibnamefont{Smy}} \bibnamefont{et~al.}
  (\bibinfo{collaboration}{Super-Kamiokande}), \bibinfo{journal}{Phys. Rev.}
  \textbf{\bibinfo{volume}{D69}}, \bibinfo{pages}{011104}
  (\bibinfo{year}{2004}), \eprint{hep-ex/0309011}.

\bibitem[{\citenamefont{Shiozawa et~al.}(1998)}]{shiozawa:1998si}
\bibinfo{author}{\bibfnamefont{M.}~\bibnamefont{Shiozawa}} \bibnamefont{et~al.}
  (\bibinfo{collaboration}{Super-Kamiokande}), \bibinfo{journal}{Phys. Rev.
  Lett.} \textbf{\bibinfo{volume}{81}}, \bibinfo{pages}{3319}
  (\bibinfo{year}{1998}), \eprint{hep-ex/9806014}.

\bibitem[{\citenamefont{Hayato et~al.}(1999)}]{hayato:1999az}
\bibinfo{author}{\bibfnamefont{Y.}~\bibnamefont{Hayato}} \bibnamefont{et~al.}
  (\bibinfo{collaboration}{Super-Kamiokande}), \bibinfo{journal}{Phys. Rev.
  Lett.} \textbf{\bibinfo{volume}{83}}, \bibinfo{pages}{1529}
  (\bibinfo{year}{1999}), \eprint{hep-ex/9904020}.

\bibitem[{\citenamefont{Kobayashi et~al.}(2005)}]{kobayashi:2005pe}
\bibinfo{author}{\bibfnamefont{K.}~\bibnamefont{Kobayashi}}
  \bibnamefont{et~al.} (\bibinfo{collaboration}{Super-Kamiokande}),
  \bibinfo{journal}{Phys. Rev.} \textbf{\bibinfo{volume}{D72}},
  \bibinfo{pages}{052007} (\bibinfo{year}{2005}), \eprint{hep-ex/0502026}.

\bibitem[{\citenamefont{Takenaga et~al.}(2007)}]{Takenaga:2006nr}
\bibinfo{author}{\bibfnamefont{Y.}~\bibnamefont{Takenaga}} \bibnamefont{et~al.}
  (\bibinfo{collaboration}{Super-Kamiokande}), \bibinfo{journal}{Phys. Lett.}
  \textbf{\bibinfo{volume}{B647}}, \bibinfo{pages}{18} (\bibinfo{year}{2007}),
  \eprint{hep-ex/0608057}.

\bibitem[{\citenamefont{Fukuda et~al.}(2002{\natexlab{b}})}]{fukuda:2002nf}
\bibinfo{author}{\bibfnamefont{S.}~\bibnamefont{Fukuda}} \bibnamefont{et~al.}
  (\bibinfo{collaboration}{Super-Kamiokande}), \bibinfo{journal}{Astrophys. J.}
  \textbf{\bibinfo{volume}{578}}, \bibinfo{pages}{317}
  (\bibinfo{year}{2002}{\natexlab{b}}), \eprint{astro-ph/0205304}.

\bibitem[{\citenamefont{Ikeda et~al.}(2007)}]{Ikeda:2007sa}
\bibinfo{author}{\bibfnamefont{M.}~\bibnamefont{Ikeda}} \bibnamefont{et~al.}
  (\bibinfo{collaboration}{Super-Kamiokande}), \bibinfo{journal}{Astrophys. J.}
  \textbf{\bibinfo{volume}{669}}, \bibinfo{pages}{519} (\bibinfo{year}{2007}),
  \eprint{arXiv:0706.2283 [astro-ph]}.

\bibitem[{\citenamefont{Malek et~al.}(2003)}]{malek:2002ns}
\bibinfo{author}{\bibfnamefont{M.}~\bibnamefont{Malek}} \bibnamefont{et~al.}
  (\bibinfo{collaboration}{Super-Kamiokande}), \bibinfo{journal}{Phys. Rev.
  Lett.} \textbf{\bibinfo{volume}{90}}, \bibinfo{pages}{061101}
  (\bibinfo{year}{2003}), \eprint{hep-ex/0209028}.

\bibitem[{\citenamefont{Desai et~al.}(2004)}]{desai:2004pq}
\bibinfo{author}{\bibfnamefont{S.}~\bibnamefont{Desai}} \bibnamefont{et~al.}
  (\bibinfo{collaboration}{Super-Kamiokande}), \bibinfo{journal}{Phys. Rev.}
  \textbf{\bibinfo{volume}{D70}}, \bibinfo{pages}{083523}
  (\bibinfo{year}{2004}), \eprint{hep-ex/0404025}.

\bibitem[{\citenamefont{Guillian et~al.}(2007)}]{Guillian:2005wp}
\bibinfo{author}{\bibfnamefont{G.}~\bibnamefont{Guillian}} \bibnamefont{et~al.}
  (\bibinfo{collaboration}{Super-Kamiokande}), \bibinfo{journal}{Phys. Rev.}
  \textbf{\bibinfo{volume}{D75}}, \bibinfo{pages}{062003}
  (\bibinfo{year}{2007}), \eprint{astro-ph/0508468}.

\bibitem[{\citenamefont{Abe et~al.}(2006{\natexlab{b}})}]{Abe:2006at}
\bibinfo{author}{\bibfnamefont{K.}~\bibnamefont{Abe}} \bibnamefont{et~al.},
  \bibinfo{journal}{Astrophys. J.} \textbf{\bibinfo{volume}{652}},
  \bibinfo{pages}{198} (\bibinfo{year}{2006}{\natexlab{b}}),
  \eprint{astro-ph/0606413}.

\bibitem[{\citenamefont{Swanson et~al.}(2006)}]{Swanson:2006gm}
\bibinfo{author}{\bibfnamefont{M.~E.~C.} \bibnamefont{Swanson}}
  \bibnamefont{et~al.} (\bibinfo{collaboration}{Super-Kamiokande}),
  \bibinfo{journal}{Astrophys. J.} \textbf{\bibinfo{volume}{652}},
  \bibinfo{pages}{206} (\bibinfo{year}{2006}), \eprint{astro-ph/0606126}.

\end{thebibliography}

\end{document}